\documentclass{amsart}
\usepackage{geometry}
\usepackage{amsthm,amsmath,amssymb,amsxtra,latexsym,mathrsfs,url}
\def\pic{\includegraphics}
\usepackage{platon,ab,weyl}
\usepackage{verbatim} 
\usepackage{cite}

\usepackage{epigraph} 
\usepackage{listings} 
\usepackage{csquotes}

\usepackage{graphicx} 
\usepackage{tikz-cd} 
\usepackage{wrapfig}

\begin{document}
\title
[Simulations to Analyze Cellular Voting Systems]{Simulations to Analyze Cellular Voting Systems for Side~Effects of Democratic~Redistricting} 
\author{
Lucius T. Schoenbaum
}
\date{\today}

\begin{abstract}
Motivated by the problem of partisan gerrymandering, we introduce an electoral system for a representative democracy called {\em democratic cellular voting} designed to make modern packing and cracking strategies irrelevant by allowing districts to be influenced directly by voters through elections. We introduce an example of a democratic cellular voting system, called CV0, that is suitable for dynamic modelling. 
We develop a modification of the theory of discrete Markov chains 
using the algebraic structure of the semiring $[0,\infty]$, 
which is used as a space of correlation coefficients. 
We use this to measure voter preferences and 
model representatives, voters, and districts in computationally feasible models with a guarantee of long-term stability. 

NOTE: this is a preliminary version of this paper. The results of the simulations are still pending. 
\end{abstract}

\maketitle

\setcounter{tocdepth}{1}

\tableofcontents

\newpage

\epigraph{
The genius of Republican liberty, seems to demand on one side, not only that all power should be derived from the people; but, that those entrusted with it should be kept in dependence on the people, by a short duration of their appointments; and, that, even during this short period, the trust should be placed not in a few, but in a number of hands.}{
James Madison \cite{FedPap37}} 

\epigraph{My intellect is so limited, Lord, that I can only trust in you to preserve me as I should be.}{
Flannery O'Conner \cite{APrayerJournal}}

\section{Introduction}\label{s.i}

Democratic norms and institutions are currently under stress \cite{CooleyDemNorms} due in part to changes in the political landscape brought about by technology used in elections \cite{SwaddleTechElections}, the media \cite{PavlikTechJournalism}, and daily life \cite{LoaderMercea2011}. 
These trends have been accompanied by remarkable trends in the United States Congress. 
Partisanship in the U.S. House is at an historic high after increasing exponentially over the past 60 years \cite{AndrisEtal}. 
Recent Congresses, under both political parties, 
have seen approval ratings drop to historically unprecedented lows, as low as 9\% in 2013 \cite{GallupApproval2013, GallupApproval2017}, 
while Gallup polling on satisfaction with the direction the country has not risen above 37\% since 2005 \cite{GallupSatisfaction}. 
A recent study has indicated that Congress does not respond to average citizen preferences, 
but does respond to preferences expressed by economic elites \cite{GilensPage1}. 
In the 2012 election for U.S. House seats, the Democrats nationally won 59,214,910  
votes, while Republicans won 57,622,827,  
according to the tally of the Clerk of the House \cite{ElectionStatsUSHouse}, yet Republicans won 33 more seats than Democrats.\footnote{Republicans won a majority by vote and by seats in 2014 and 2016. Prior to 2012, the party whose national vote majority exceeded the other major party won the majority going back until 2000, the last year that elections were held based on the 1990 Census. In that year, Republicans lost 2 seats while Democrats won 1. The vote differential nation-wide was R/D 410,216.} There is evidence \cite{BrennanExtremeMaps,StephanopoulousMcGhee} that this imbalance occured due to partisan gerrymandering in a handful of states after the 2010 census. 

Attention to the issue of gerrymandering in the research community has grown in recent years \cite{AlexeevMixon,BuzasWarrington,GilliganMatsusaka,McCartyPooleRosenthal}. 
In \cite{BernsteinDuchin}, Bernstein and Duchin define gerrymandering as the drawing of political boundaries with an ulterior motive. 
Gerrymandering can be {\em racial gerrymandering,} which seeks to minimize the influence of racial minority groups in Congress, {\em incumbent gerrymandering,} in which politicians ensure victory for incumbent politicians regardless of their party affiliation, or {\em partisan gerrymandering,} in which districts are drawn that artificially inflate the number of members from one political party. Gerrymandering is achieved by creating useful votes for one group and wasted votes for a competing group. This can be 
by {\em packing,} or drawing a district that contains an overwhelming majority of voters from one party (who therefore waste the surplus of votes), or by {\em cracking,} or dividing a geographic region that favors one outcome into two or more smaller ones inside of districts containing a majority population that favors a different outcome. 

The suspicion or accusation of gerrymandering has existed at least since the term was first coined 
by the {\em Boston Gazette} in 1812. 
The Supreme Court has weighed in on the issue 
three times. 
In earlier eras discussion in the courts has been over whether 
gerrymandering is a justicial issue. Today this is settled: 
intentional partisan gerrymandering is a justicial issue. However, there is no agreement at all in the court system on what constitutes intentional gerrymandering, and if so the basis on which it can be prosecuted. 
Since {\em Davis v. Bandemer} (1986), the 
view of the Supreme Court has been 
that the mere lack of proportional representation 
is not a sufficient basis for unconstitutional discrimination. 

The {\em efficiency gap}, introduced 
by law professor Nicholas Stephanopoulous and political scientist Eric McGhee 
\cite{StephanopoulousMcGhee}, is defined as the number of votes wasted divided by the number 
of votes cast. Stephanopoulous and McGhee argue in part that the efficiency gap 
isolates partisan gerrymandering's effect, and provides an accurate measure of its magnitude that 
did not exist when the {\em Davis v. Bandemer} decision was made. 
An unconstitutional state map according to Stephanopoulous and McGhee's evaluation 
would have an efficiency gap higher than 7-8\% in favor of the party in control. 
In arguments before the Supreme Court in {\em Ghil v. Whitford}, heard on October 3, 2017, 
districts drawn for the Wisconsin State Assembly 
by the Republicans were held to be unconstitutional based on evidence of intentional partisan gerrymandering, along with the fact that the efficiency gaps were determined to be 10\% in 2012 and 13\% in 2014. 
The Supreme Court will decide among other things whether the efficiency gap can be used as a basis for rulings on the constitutionality of district maps drawn by one party. 

This paper approaches the issue of gerrymandering through mathematical modeling of voting systems other than single-district, plurality voting, i.e., our familiar voting system of today. The basic point of departure is a voting system called {\em democratic cellular voting} that specifically targets the issue of gerrymandering. 
Rather than developing a framework for thinking about democratic cellular voting as a general family of voting systems, we invest the majority of our time in what is already rather a challenge, namely defining an example system, called CV0. 

The basic ideas in this first system CV0, as described in more detail in sections \ref{s.cells} and \ref{s.cv0}, are the following:
\enu
	\item cellular partitioning as a foundation for district creation, resembling the finite element method \cite{ReddyFiniteElement}, a common approach to systems in continuous domains, 
	\item overlapping districts obtained by simple weighted statistical averaging and 
the ``one person, one vote'' principle applied as a system axiom, 
	\item selection of districts through voter and their respective candidate 
decision-making and preference-weighting under competitive cooperation and incentivization regimes,
	\item geometric compactness metrics 
to prevent the incursion of political bias and gerrymandering in the drawing of cells, 
	\item a notion of valence revision (of a representative) that enhances voter accountability even for representatives in ``safe'' districts.
\Enu
The main point is that the combination of compactness at the cellular level and democratic input at all higher levels 
throws up a bulwark against gerrymandering (see section \ref{ss.gerry}). 

In sections \ref{s.math} and \ref{s.math2} we develop the mathematics we need for CV0's modeling. 
In section \ref{s.cells} we describe a democratic cellular voting system in skeletal outline 
as though it were a quote-unquote ``real'' voting system. 
This will help us with modeling, in particular identifying which parts of the model are realistic and which parts are ideal. 
In section \ref{s.cv0} we describe the model CV0 itself in broad outline. We conclude with final remarks and mention how the work might be ongoing in section \ref{s.conc}. 

This is a preliminary version of this paper. The results of the simulations are still pending. 
We will report on these simulations when they are completed, in an update to this paper.

\section{The Semiring $[0,\infty]$}\label{s.math}

Semirings have recently received attention in tropical geometry, 
where much focus is on the max-plus algebra $\real \cup \set{-\infty}$. 
Here, our focus will instead be on the algebra $[0,\infty]$ of nonnegative numbers, plus an 
ideal infinite number. 
Let us observe that under addition and multiplication, $[0,\infty]$ is a semiring with 
not one, but two zero elements $Z$ satifying $a \cdot Z = Z \cdot a = Z$ for all $a$, 
with one of them dominant in the sense that $Z_1 + Z_2 = Z_1$. 
Moreover, just like the max-plus algebra it is a semifield, provided we exclude the inverse requirement for both of these 
two zeros. 

We consider the action of $[0,\infty]$ on state vectors, that is, vectors of real numbers 
in the interval $[0,1]$, given by 
\begin{equation}\label{e.expoaction}
a \cdot x = x_i ^a
\end{equation}
where $a \in [0, \infty]$, $x$ is a state vector, $x_i$ is the $i$th component of $x$, and we put $a^\infty = 0$ for all 
$a$ in $[0,1]$. This includes $a = 1$, which is the symmetric opposite of $0^0 = 1$. 
With respect to this action, two zero elements $0$ and $\infty$ can be thought of probabilistically as being overrides, 
making an event either certain or impossible, respectively. 

We assemble state vectors (assume the size of the vectors to be fixed) into a semimodule $\sV$ over $[0,\infty]$, which 
we call a semi-vector space, under the operation 
$$x \oplus y = (x_i  y_i)$$
and the action (\ref{e.expoaction}) of exponentiation. 
Observe that there is an additive unit $(1,1,\dots,1)$ and an additive zero $(0,0,\dots,0)$ in $\sV$. 
The rest of the axioms 
are easy to verify. 

Linear maps on $\sV$ may now be defined. 
If we define the natural inner product 
$$(x,y) = y_1^{x_1} y_2^{x_2} \dots y_n^{x_n}$$
and matrix multiplication based on this 
inner product, we obtain a nonassociative ring of matrices. 
This ring may be thought of as loglinear, not 
in the statistical sense \cite{Agresti1}, but in the sense that a linear map described by a matrix 
$$f(x) = A \cdot x,$$
where $(\cdot)$ is the extension of the action (\ref{e.expoaction}), 
can be computed using the formula
$$f(x) = \exp(A \log x), \quad x \in \sV,$$
where $A$ is a matrix over $[0,\infty]$ and the functions $\exp$, $\log$ are taken component-wise. 
Note the appearance of ordinary matrix multiplication, which extends to compositions: if
$$g(f(x)) = B\cdot A \cdot x,$$
associating from the right, then we have this equal to the expression
\begin{equation}\label{e.loglinear}
\exp(BA\log x),
\end{equation}
where the product $BA$ is matrix multiplication. In fact, this 
relationship to matrix multiplication arises as a substitute for associativity 
$$A\cdot (B \cdot C) = (AB) \cdot C,$$
as the multiplication ($\cdot$) is not associative. 
This reasoning clearly extends to arbitrary products, giving a feasible way of computing 
such matrices. 
The semi-vector space $\sV$ can be extended to a full vector space, and norms 
for these spaces can be defined. 

%We will use these observations as a mathematical basis for a three-valued voting system in which votes can be positive, negative, or neutral. 
%In section \ref{s.math2} we will extend this for modeling purposes, and by modifying the theory of discrete Markov chains, 
%derive guarantees about long term behavior of models of voting populations. 

\section{Steady State Theory for Models with Causation and Correlation}\label{s.math2}

\subsection{The desired modification of discrete Markov chain theory}\label{ss.markov}

Our first goal in this section is to generalize discrete Markov chain theory in the manner needed for our purposes. 
The main issue is that the class of matrices of interest has real 
entries. 
There has been work done on extending the Perron-Frobenius theorem and related theory to 
this class of matrices \cite{NaqviMcDonald,ZaslavskyTam}, but it is focused on the eventually nonnegative case, 
while we have in mind the semiring $[0,\infty]$. 
So our attention is on the quasi-stochastic case (defined below). 
Such matrices need not be eventually nonnegative, and they may not have a positive eigenvector. 
An example of such a matrix is 
$$\left(\begin{tabular}{ccc} 0.9 & -0.1 & 0.1 \\ 0.2 & 0.9 & 0.2 \\ -0.1 & 0.2 & 0.7 \end{tabular}\right)$$
with steady state vector $(-1/6, 2/3, 1/2)$.

Some basic terminology: 
We say that a vector $x$ is {\em positive,} ({\em nonnegative}), $x > 0$ ($x \geq 0$), if every component $x_i$ satisfies $x_i > 0$ ($x_i \geq 0$, respectively). In particular, all components are real. 
A matrix $A$ is positive (nonnegative) if it is positive (nonnegative) when viewed as a vector. 
A matrix $A$ is {\em diagonally positive} if $a_{ii} > 0$ for all $i$. 
More terminology: 
A matrix $A$ is {\em eventually positive} (or {\em primitive}) if 
there exists $k$ such that $A^k$ is positive. 
$A$ is {\em eventually nonnegative} if there exists $k$ such that $A^k$ is nonnegative. 

The {\em directed graph} of square matrix $A$ is the directed graph 
$G(A)$ with vertices labeled $1, \dots, n$, where $A$ has size $n$, 
and a path from $i$ to $j$ if and only if $a_{ij} \neq 0$. 
$A$ is {\em irreducible} if $G(A)$ is strongly connected (cf. \cite{ZaslavskyTam}). 

Recall that a matrix $A$ has {\em dominant eigenvalue} $\lambda_1$ if the eigenvalues $\lambda_1, \lambda_2, \dots, \lambda_n$ of $A$ 
(counted by multiplicity so that $n$ is the side length of $A$) satisfy 
$|\lambda_1| < |\lambda_2| \leq \dots \leq |\lambda_n|$.

All matrices we study will be weakly stochastic:

\dfn\label{d.weaklystochastic}
A matrix $A$ is {\em weakly stochastic} 
if it is square, has real entries, and has columns that sum to 1. 
\Dfn

\begin{figure}[t]
\begin{center}
\pic[width=300pt]{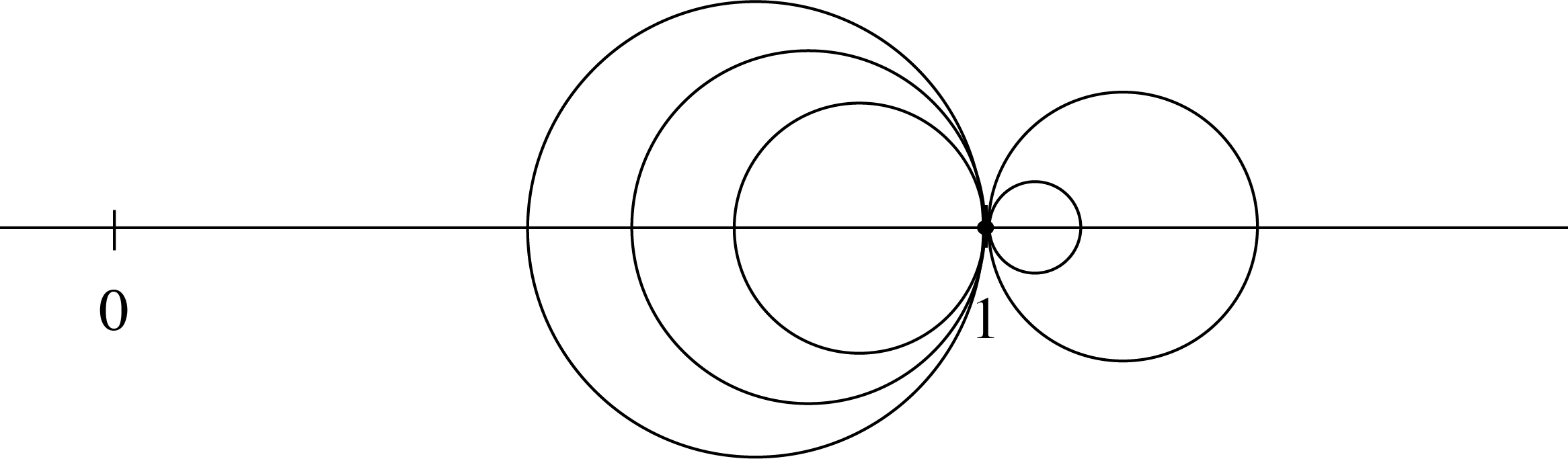}
\end{center}
\caption{}
\label{f.gershgorin}
\end{figure}

Let $A$ be weakly stochastic. Thus 
$A$ has 1 as eigenvalue, since $A$ and $A^T$ have the same eigenvalues and $A^T$ trivially has eigenvector $(1,\dots,1)$ with eigenvalue 1. 
Moreover all other eigenvalues $\lambda$ of $A$ satisfy $|\lambda| < 1$ or $|\lambda| > 1$ due to the 
Gershgorin circle theorem 
(see {\sc Figure} \ref{f.gershgorin}). 

The following terminology for describing features in a weakly stochastic matrix is helpful for our purposes. 

\dfn\label{d.inflationetc}
Let $A$ be a diagonally positive matrix. 
An {\em inflationary term} is a term $0<a_{ii} < 1$ on the diagonal. 
A {\em deflationary term} is a term $1 < a_{ii} < \infty$ on the diagonal. 
An {\em zero inflation term} is a term precisely $a_{ii} = 1$ on the diagonal. 
For present purposes, a {\em pair}, referring to a matrix $A$, is a pair $a_{ij}$ and $a_{ji}$ 
of off-diagonal terms. 
A {\em control feedback pair} is a pair with 
each member of the pair having different sign. 

A {\em resonant feedback pair} is a pair with both members negative. 
An {\em op-resonant feedback pair} is a pair with both members positive. 
\Dfn

\dfn\label{d.nonreinforcing}
A matrix $A$ is {\em non-reinforcing} if every inflationary term is not a part of a resonant feedback term, 
and every deflationary term is not part of an op-resonant feedback term. 
\Dfn

\dfn\label{d.quasistoch}
A matrix $A$ is {\em quasi-stochastic} if:
\enu
	\item it is square, 
	\item it is weakly stochastic and diagonally positive, 
	\item it is non-reinforcing. 
\Enu
\Dfn

\conj
A quasi-stochastic matrix has dominant eigenvalue 1, hence the limit 
$$\lim_{k \to \infty} A^k = \pi {\bf 1}^T$$
exists, 
where ${\bf 1} = (1,1,\dots,1)$ and $\pi$ is a vector. 
\Conj

We will prove this in an update of this paper. If it is not true, 
it will assuredly be true in enough cases not to in any way imperil the 
desired application of section \ref{s.cv0}.

\section{Voting Cells}\label{s.cells}

\subsection{Geography}\label{ss.cells}
Recall that a democratic cellular voting system is one in which districts are formed out of cells in a democratic way (i.e., through elections). 
The land is therefore separated into voting cells. 
We now give a technical definition for the cells:

\dfn\label{d.cell}
A {\em voting cell} is the basic unit of voting. It is an extent of land that is contiguous (connected) 
to the the maximum extent permitted by natural land features. 
Cells may not overlap and cells must cover the nation's entire area. 
Each voter belongs to (is a member of, etc.) the cell containing the voter's registered address for voting purposes. 
Cells must satisfy the {\em Gerrymandering Requirements} which may vary depending on how the system is implemented (see below). 
A voting cell's {\em area} is the area in the usual sense (as measured by surveyors, etc.) 
Two cells are {\em adjacent} if they share at least one boundary point in common. 
Each cell, we shall say, contains approximately 1000 voters ($\pm 10\%$) at its formation. As cells grow, they are divided into two, each of them 1000$\pm 10\%$, as before.\footnote{Another possibility is that maps are redrawn periodically into new cells that respect the Gerrymandering Requirements. However, 
if cells change during the cellular map revision, all the existing districts are potentially affected and the preference information (from both voters and representatives) reflected in the previous map gets lost. If instead the cells are allowed to divide (say, along the boundaries suggested by the splitline algorithm) then the existing multi-cellular districts can continue to persist in the new maps as before.} 
\Dfn

\begin{figure}[t]
\begin{center}
\pic[width=300pt]{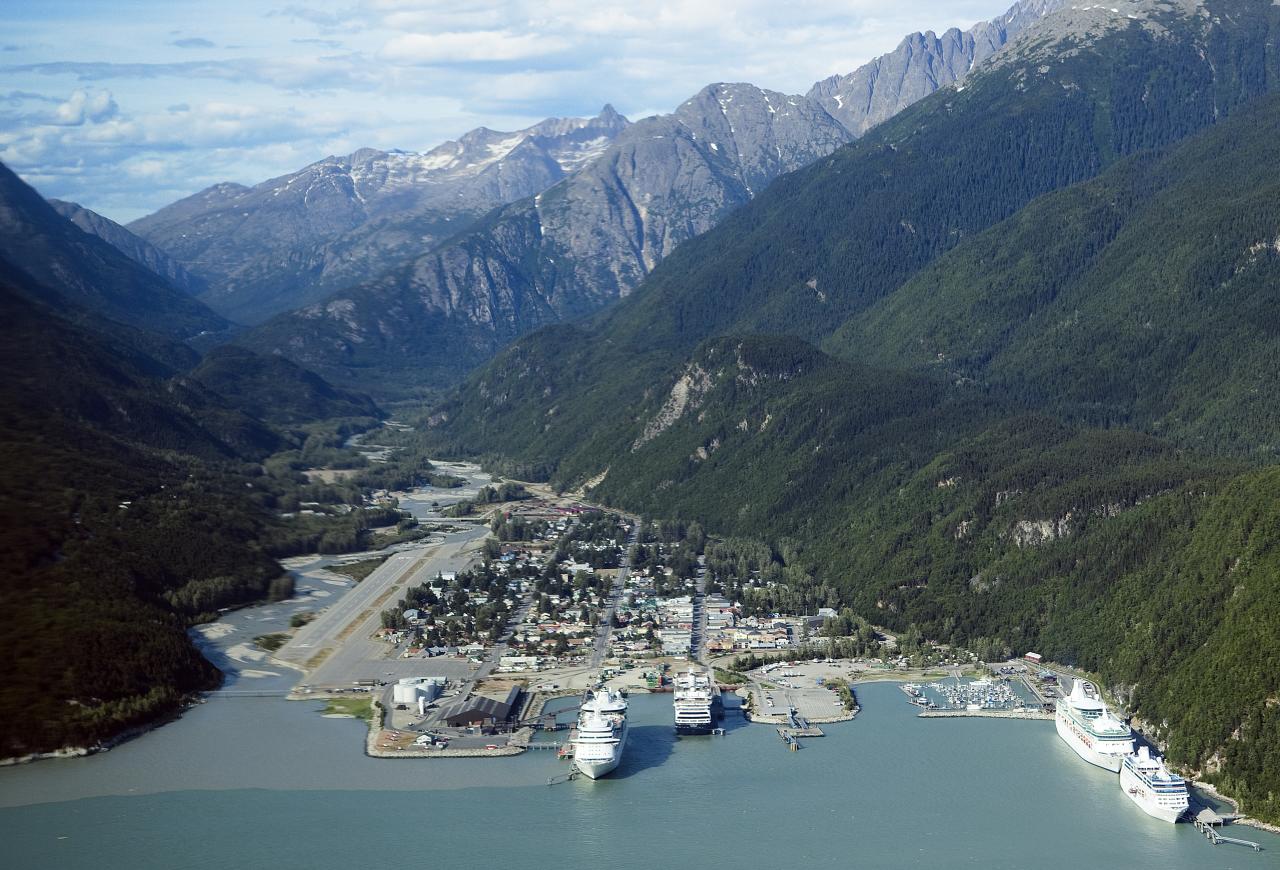}
\end{center}
\caption{Skagway, Alaska (Pop. 920) {\tiny Photo: \cite{SkagwayPhoto}}}
\label{f.skagway}
\end{figure}

How many people are 1,000 people? 
\enu
	\item The population of the Municipality and Borough of Skagway, Alaska is 920, or around 400 households (see {\sc Figure \ref{f.skagway}}). 
	\item Estimating that there are 28,000 blocks in the city of Chicago, which has a population of 2,695,598, a city block would contain on average 96 people, so 1,000 people would be, geographically, roughly 10 blocks. Chicago would be divided into roughly 2,500 cells the size of about 10 city blocks each. 
	\item Estimating that there are 2,000 blocks on the island of Manhattan (New York County) which has a population of 1,643,734, a city block would contain on average 800 people. Manhattan would be divided into roughly 1,600 cells, each the size of 1-2 city blocks each. 
	\item Estimating that an average ``small town'' has a population density around 800 people per square mile, such an area would be divided into cells that are geographically approximately 1 square mile. 
\Enu
Each of these cells would be the minimal areas which, if elected, a citizen could represent in Congress. The area of the 10th most populous state, Ohio, for example, would be separated into approximately 11,600 cells. In most states there would be (as in this example) around 700 cells per modern U.S. House district. This applies to any state, as congressional districts are all drawn to be the same size by head count, or approximately 700,000 today. 

The Gerrymandering Requirements might be one of the following:
\enu
	\item The cells might be determined by the splitline algorithm \cite{RangeVoting}. 
	\item The cells might be required to pass a strict, or area-weighted Polsby-Popper compactness threshold. Area weighting might be employed in order to relax the compactness requirement for cells with a large area. 
\Enu

Affairs in each cell 
would be managed by a Chief Cellular Officer, who is either elected or appointed by the reasonable choice of official, 
such as the governer of the state.
Chief Cellular Officers report to Regional Offices. 
The Chief Cellular Officer must keep an office, and be available for meetings with citizens. Cellular offices might be installed at a local City Hall, or inside the nearest post office.

\subsection{Valence}\label{ss.valence}
\begin{figure}[t]
\begin{center}
\pic[width=220pt]{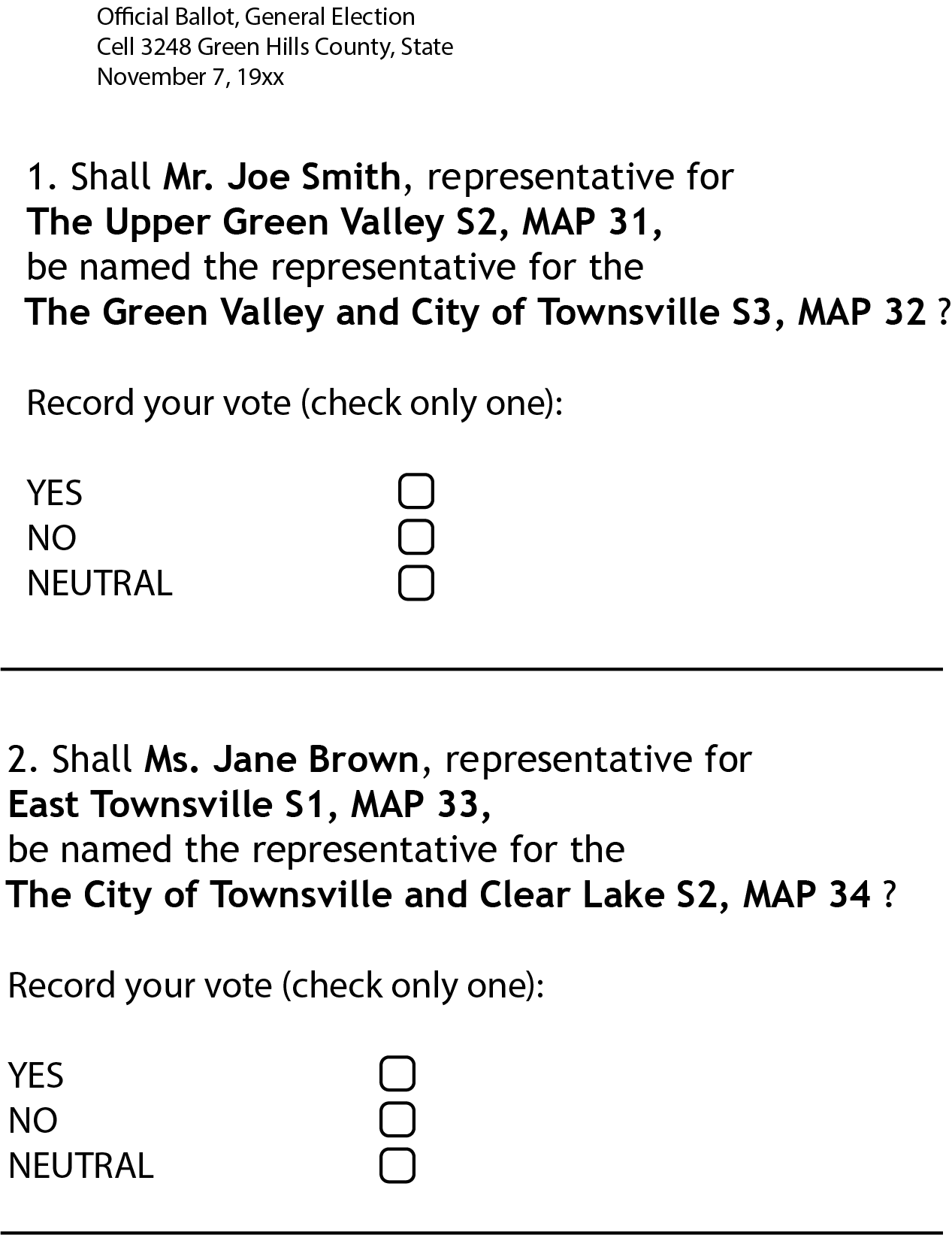}
\end{center}
\caption{Example ballot. Only two questions shown.} 
\label{f.ballotexample}
\end{figure}

The voting system we propose to model in this paper is as follows. 
Voting occurs once per {\em cycle} $t$; we do not specify the length of a cycle. 
Voters vote on {\em candidates} for office, who may be {\em incumbents}, in the sense that they represented the cell 
during the previous cycle $t-1$. 
Henceforth in this discussion we make no distinction between candidates and incumbents and 
refer to them all as {\em representatives}. 
Voters vote on representatives once per cycle, and their vote may 
be either {\em Yes}, {\em No}, or {\em Neutral}. (See {\sc Figure} \ref{f.ballotexample} 
for a simple, informal example ballot.) 
These votes are tallied to determine each representative's {\em valence}. 
Valence is a formal, ballot-box-determined version of what is known as {\em strength}, or {\em political strength}, in politics. 
The idea is simple: the more people vote for you, the more your valence goes up. 
The more people vote against you, the more your valence goes down. 
People who vote but vote neutral do not affect your valence, {\em but,} 
valence only depends on the size of the voting population, not the population size at large. 
With that being said, here is the formal definition of valence. 

Let $C = c_1, \dots, c_K$ be the set of cells, 
let $R = r_1, \dots, r_M$ be the set of representatives. 
During the cycle $t$ each cell has a set $P(c,t)$ of active voters. 
For each representative $r$, for each cell $c$, and for each cycle $t$, the value 
$$A(r, c, t) = \sum_{\substack{v \text{ votes Yes} \\ v \in P(c,t)}} 1 \quad - \sum_{\substack{v \text{ votes No} \\ v \in P(c,t)}} 1$$
is the number of Yes votes minus the number of No votes for representative $r$ in cell $c$ in cycle $t$. 
Then the {\em raw tally} score is: 
$$Tally(r,c,t) = \frac{A(r,c,t)}{|P(c,t)|} \cross 100.$$
Here's a simpler breakdown that a member of cell $c$ might use: 
if $Y$ is the number of Yes votes for candidate $r$, and $N$ is the number of No votes for candidate $r$, and $NT$ is the number of Neutral votes for candidate $r$, then 
$$\frac{Y - N}{Y + N + NT} \cross 100$$
is the raw tally awarded to representative $r$ by cell $c$. Notice that the quantity $Y + N + NT$, which we call the {\em raw valence} of the cell $c$, 
does not depend on representative $r$ 
and is equal to the number of members of the total population of cell $c$ who voted in cycle $t$. 
Raw tallies thus fall between $-100$ and $100$. 
The raw tally is therefore a measure of simply who won and who lost in the cell, and by how much. 

Based on the raw tallies, each representative $r$ is awarded a {\em valence} by cell $c$ in cycle $t$, which holds during cycle $t+1$: 
$$Val(r,c,t+1) = \max \left(0, \frac{Tally(r,c,t)}{\sum_{r \in R, Tally(r,c,t) > 0} Tally(r,c,t)}\right) \cross 100.$$
Note the invariant that for all cycles $t$,
$$\sum_{r \in R} Val(r,c,t) = 100.$$
Representatives whose valence at the cellular level is zero has that cell {\em flagged for removal} from their district (the district they are running in, 
not the district they hold). 
In other words, they are deemed to have ``lost'' in that cell and cannot represent that cell in Congress. 
Such a cell will not be included in that representative's district during the next cycle $t+1$. 

Now let $P(t)$ be the set of all voters who voted in all the cells in cycle $t$. 
We define the {\em valence} of representative $r$ during the cycle $t+1$ to be:
$$Val(r,t+1) = \frac{ \sum_{c \in C} |P(c,t)| \cdot Val(r,c,t+1) }{|P(t)|}.$$
We have thus assigned each representative a valence between 0 and 100. 
We have defined the total invariantly in the sense that for every cycle $t$, 
$$\sum_{r \in R} Val(r,t) = 100.$$
Note that the contribution to $Val(r,t+1)$ made by cell $c$ in cycle $t$ depends on the quantity $|P(c,t)|$. It will be higher if more people vote. 

Votes in Congress during cycle $t$ are taken as follows: 
for any vote on a motion to proceed, on a bill's passage, etc., a vote is made by each representative, who votes 
either Yay, Nay, or Abstain. 
The clerk records the Yay votes and the Nay votes, and notes who has abstained. The vote is then calculated as:
$$R = \sum_{r \text{ votes Yay}} Val(r,t) - \sum_{r \text{ votes Nay}} Val(r,t).$$
If $R > 0$, the motion is passed. In a tie, the vote may be retaken.

\subsection{Cognitive Burden}\label{ss.cognitiveburden} 
There is a problem here that we do not wish to disguise. 
It has to do with what we call {\em cognitive burden,} 
or the work that voters are expected to do in order to make the voting system function. 
For our work in this paper, 
we have chosen to leave this problem out of consideration. 

For some short discussion of where we stand on the cognitive burden issue, see section \ref{s.conc}.

\subsection{The Question on the Ballot}\label{ss.question}
We can easily imagine a more refined measure of voter preference in the 
tally of the election. 
For instance, we could ask voters to vote on a scale of 1 to 5, with 3 being neutral, 4 being 
a vote of support, 5 being strong support, 2 for do not support, and 1 for strongly do not support. 
We could imagine going even further. 
We could imagine asking voters to quote a number between 0 and 1! 
However, some subset of the voters would be expected to recognize that in order to have their 
own vote weigh most heavily on the outcome, 
they should adopt the strategy of voting as strongly as they are allowed to, 
in every instance where they have an opinion one way or the other, either Yes or No. 
Other voters, who might not be as inclined to abuse the system, 
would recognize that other voters might adopt this strategy, 
and respond in kind. 
Any such system can therefore be expected to have at most three options for the 
voter to choose from on any question on the ballot: Yes, No, or Neutral. 

\subsection{Political Parties and the Problem of Career Advancement}\label{ss.part}
It is historically rare for a member in the U.S. House of Representatives, the lower house of Congress, 
to resist the will of one of the major parties \cite{HouseArchivesPartyDiv}. 
There are many reasons for this, one of which is that for a House member, there is a remarkably small window for professional advancement. 

House members, like all politicians, having begun a career as a statesman or stateswoman serving their country, naturally aspire to run for a higher office. 
What can they run for? House members sometimes run for president. Since 1900, 
the former Presidents who are also former House members 
are: George H. W. Bush (Texas's 7th district), Richard M. Nixon (California's 12th district), and John F. Kennedy (Massachusetts's 11th district). These former representatives all ran successfully for president---but in these instances, the jump from House member to president was indirect (Senator~seats, in the case of Nixon and Kennedy, and Vice~President, in the case of Nixon and Bush). Other former House members who were presidents (Gerald R. Ford and Lyndon B. Johnson) gained the office without having run in an election. 

Another option is to run for the Senate. In 2004, for example, David Vitter successfully ran for Senate, having represented Louisiana's 1st district in the U.S. House of Representatives from 1999 to 2002. Tom Cotton did so successfully in 2014, having represented Arkansas's 4th congressional district since 2013. 
There are innumerable similar cases. 
In 17 instances all prior to the modern era (post World War II), House members were also appointed to the Supreme Court. 

House members also sometimes choose to leave Washington, DC to run for high office in their home state. 
This occurred for example in 2010 when Rahm Emanuel, a former member of the U.S. House of Representatives for Illinois's 5th congressional district, left Washington, DC to run for mayor of Chicago, the office he presently holds. In another instance, Bobby Jindal of Louisiana, a former member of the House for Louisiana's 1st congressional district, returned to his home state to run successfully for governor in 2007, the office he held from 2008 to 2016. 

To an outsider, it might 
seem cause to wonder why 
it is a fairly regular occurence that a statesman or stateswoman leaves the nation's capital, after having won a position in the nation's foremost lawmaking body, 
and instead opts for life in a state office. 
It is sometimes said that this is because as a U.S. House member, you are but one voice in a body of over 400 individuals. 
This may weigh on some decisions. 
But the simple math that House members face also plays a role. 
By the Permanent Apportionment Act of 1929, the size of the U.S. House of Representatives is fixed at 435 (leading to an average district population size of approximately 700,000 today). 
With governorships included, this gives a body of 435 room for advancement to only 151 new seats, or 152, if you include Vice President. 
Then again, the vast majority of those seats are not available to all representatives, because they funnel through the states. In a state such as Ohio, for example, there are 16 House members having 4 seats for advancement between them. 
Two of these come up for reelection only once every six years, and are not term limited offices; these are the two Senate seats. 
One of these is the presidency; the other is the governorship. 
The smaller a state's population, the greater the opportunity for advancement: Alaska, Delaware, Vermont, Wyoming, Montana and the Dakotas all have the smallest representation possible (one~member). These members each have 4 paths for advancement. 
Their opposites are the four states with 20 or more representatives: California (53), Texas (36), New York (27), and Florida (27). 
California, with a delegation of 53 House members, also has 4 seats for advancement, one of which is the same presidency that politicians from all states aspire to. 

No matter how popular a representative becomes, there is no way he or she can grow her base (the population 
of his or her district) except by one of these crowded paths. 
This inertia of a House seat applies not only if one attempts to leave it, but also if it is held, in that there is no incentive to increase enthusiasm or improve one's popularity in one's district above and beyond the plurality, first-past-the-post (in most states) vote that occurs once every two years---often, today, in a safe seat. 
Once you are elected, being a House member is an environment lacking natural opportunities for long term advancement, with one notable exception: the opportunities on offer 
through the party system. The party you belong to provides you with the opportunity to rise up the ranks, and thereby obtain powerful committee seats and leadership roles, have an influence on scheduling, bill markups, appointments, and so on. 
Thus the limited avenues for advancement 
create the conditions for high, even obsequious loyalty to party, and voting that goes down mechanically along party lines. 
The phenomenon that is expected under these incentives is documented by Ansolabehere, et al. \cite{AnsolabehereEtal}, who note that, 
in their words:
\blockquote{All candidates, the thinking goes, pursue centrist strategies within their districts, so the winners should accurately reflect the desires of the greatest number of voters. Evidence for this belief is the strong correlation between the winners' (incumbents) roll-call voting records and the ideological leanings of the districts. Our analysis suggests this pattern is much weaker than commonly believed. While there is a statistically significant amount of responsiveness, it is not the main story. Competing candidates in congressional elections almost never converge. Instead, the strong correlation between incumbents' and districts' ideologies arises almost entirely because voters are presented with largely partisan choices and select the candidate whose party more closely resembles them \cite[p. 152]{AnsolabehereEtal}.
}

The career outlook of a representative in a democratic cellular voting system looks much different. 
A successful representative can choose to target an enlarged base with every new election, and tailor that base to their own personal strengths and weaknesses as a candidate. 
Representatives can contemplate expanding into neighboring state-sized areas and, if successful, eventually ride to power in entire regions and parts of the country. 
There is fine-grained, direct voter accountability to 
countervail the influence of a party or other interceding organization. 
Members can appeal to voters for their support, or denial thereof, if they would like to operate as an independent or as a party loyalist. 
If a popular decision is made against a party line, for example, 
or if an unpopular decision is made with it, 
voters can express their thanks, as it were, to the representative by an increase of his or her valence (see section \ref{ss.valence}), and conversely express their disapproval in a way that would be felt by the representative, no matter whether he or she won or not in the next election. 
This is true regardless of whether the representative is in a safe seat.

%The political party system, as commonly understood, is a very complex topic warranting further investigation within the context of democratic cellular voting systems. 
%However, we wished to call attention to the problem of career advancement, which played an important role in the thought that led to the system we have outlined. 

\subsection{Gerrymandering and District Cultivation}\label{ss.gerry} 
We can now survey the bulwark against gerrymandering mentioned in the introduction, put up by a democratic cellular voting system featuring the notion of valence of section \ref{ss.valence}. 
The rolling revision of valence creates an incentive to hew closely to the majority will of a constituency, even one that exceeds a simple majority. 
This eliminates the problem of packing, 
because a well supported candidate will have the valence to show for it, compared to one that scrapes past the post: no votes wasted; every vote carries an iota of valence for every representative sent to Congress. 
What about cracking? 
Because all valences are strictly positive, and percolate up through the results in the cells, 
voters (at least, up to the cellular level) have a say on whether a candidate can represent them in congressional votes. 
No population can be diluted against its will through any strategic maneuvering by a representative. 
Should it lack the representation that it cries out for, a concerted cell can take matters into its own hands. Members can recruit one of their own, effectively taking their valence off the table. 
Hence there need be no wasted votes by cracking either. The efficiency gap, if considered conceptually as a measure of wasted votes, should remain at a very low level through time. 

With all the necessary caveats that the approach is very new and untested, 
democratic cellular voting systems therefore offer a potential solution to the problem of gerrymandering. 
Consider for example the impossibility theorem of Alexeev and Mixon \cite{AlexeevMixon}, showing that it is not always mathematically possible (!) to satisfy the efficiency gap and Polsby-Popper compactness requirements simultaneously. 
According to legal arguments that go along somewhat similar lines, courts have sometimes argued (on the basis that it is ``always difficult to prove a negative'') that the legal system cannot act based on social science research \cite{RyanDeseg} such as the efficiency gap metric because, among other issues, actors's intentions get lost in the process of statistical aggregation. 
We currently await the court's decision in {\em Gil v. Whitford}. 
Whatever happens, however, gerrymandering is an issue that is handled differently, and well, by the democratic cellular voting model. While districts might take on ``gerrymander'' shapes, it will not be due to the effects of gerrymandering, but through the process of voters finding representatives that speak to them, and through them to the legislature. 
These mathematical and legal clouds therefore no longer haunt us. 

However, the cure's side effects are unknown.

\section{Dynamic Models of Voter Populations and CV0}\label{s.cv0}

In this section we use the mathematical framework of sections \ref{s.math} and \ref{s.math2} to dynamically 
model voters and representatives, as they create districts that evolve 
on rules that follow the outline of a democratic cellular voting system 
in section \ref{s.cells}. 

When we model voters and representatives, and allow 
the simulation to evolve over time, 
we wish to monitor whether it drifts in an unstable direction, due to polarization or some other type of clustering. 
Therefore we seek a guarantee that if the system drifts in a unstable direction, it is due to the effect of the voting system and 
not the way in which the model 
is configured. 
For this, we use the modified Markov chain theory of section \ref{ss.markov} 
to construct a model with a continuous 
state space that comes with a guarantee of long term behavior, i.e., a steady state. 
For the purpose of such a model, the mathematical notions of the previous section 
apply to correlation coefficients
on parameters. 
In the terminology of section \ref{ss.markov}, 
a control feedback pair $(+,-)$ is like two parameters, such as eating and napping. 
If I eat a meal, then my chance of taking a nap goes up. But if I take a nap, my chance of eating a meal goes down. 
An op-resonant feedback pair $(+,+)$ is like losing a job and developing a bad habit. 
If I lose a job, my chance of developing a bad habit goes up. If I develop a bad habit, my chance of losing a job goes up. 
A resonant feedback pair $(-,-)$ is like having money and investing wisely. 
If I have money, my chance of investing it wisely increases. If I invest money wisely, my chance of having money increases. 
These pairs are used to define correlation and causation weights inside the CV0 model. 
For example, a voter's attribute of employment status might be correlated to their likelihood of moving, 
and this weighting might be causal in the sense that one might trigger the other, but the other might 
not be expected to trigger the first. 
Tuning the model is a matter of setting values in a quasi-stochastic matrix. 

Note that the steady state configuration will not necessarily be obtained in the model over time, due to the effects of noise, voting, and relocation. 

\subsection{Assumptions of the model}\label{ss.modelassumptions}
Our model has voters, cells, and representatives. Here are some basic assumptions: 
\enu
	\item A system or set of {\em attributes} is assigned to each model element (voter, cell, or representative). For a voter these could be, for example: health, 
		job status, cost of living, income. For a cell, these could be for example: natural beauty, natural resources, population density. 
		For a representative, these could be for example: talent, skill, experience, charisma, respect among colleagues, and so on. 
	\item An amount of conserved raw material 
		exists for distribution to each element (voter, cell, or representative) 
		in the slots we use to track attributes over time. 
		This replaces the assumption in the theory of discrete Markov chains that one works with probability vectors. 
		This allows the dominant eigenspace to be normalized; otherwise the system's long term behavior 
		becomes sensitive to initial conditions. 
		In the case of voters, this is handled by distributing age, as a hidden variable, to the other parameters. 
		In the case of cells, this is handled by setting prosperity and comfort slightly in opposition. 
		In the case of representatives (who are subject to retirement), this is handled by being assigned at random. 
	\item Attributes can be correlated (each boosts the other), or causally connected: one can have a direct {\em effect} on the other. 
		It is theoretically possible for one to affect the other positively, while the other in turn affects the first negatively: 
		it is all coming down to two values for each pair of attributes chosen. 
\Enu
We make a few notes about behavior to expect in models using the $[0,\infty]$ approach:
\enu
	\item ``Too much of a good thing'' principle: Any item $A$ 
that has a causal connection to another item $B$ 
will have effects which taper off as $A$ is augmented. 
	\item ``No grounding'' principle: The meaningful regime for any item $A$ will not include values past a certain small amount, and the system arrives, when it is defined, with a guarantee that every item $A$ will regress back to higher values when it dips. \label{i.ground} 
This is because the way the system is set up, in order to allow for a guarantee about long term behavior (steady state), an item 
correlated to another item will behave strangely if its value is very small (close to zero). 
\Enu

\subsection{A typical timestep}\label{ss.timestep}

\begin{figure}[t]
\begin{center}
\pic[width=400pt]{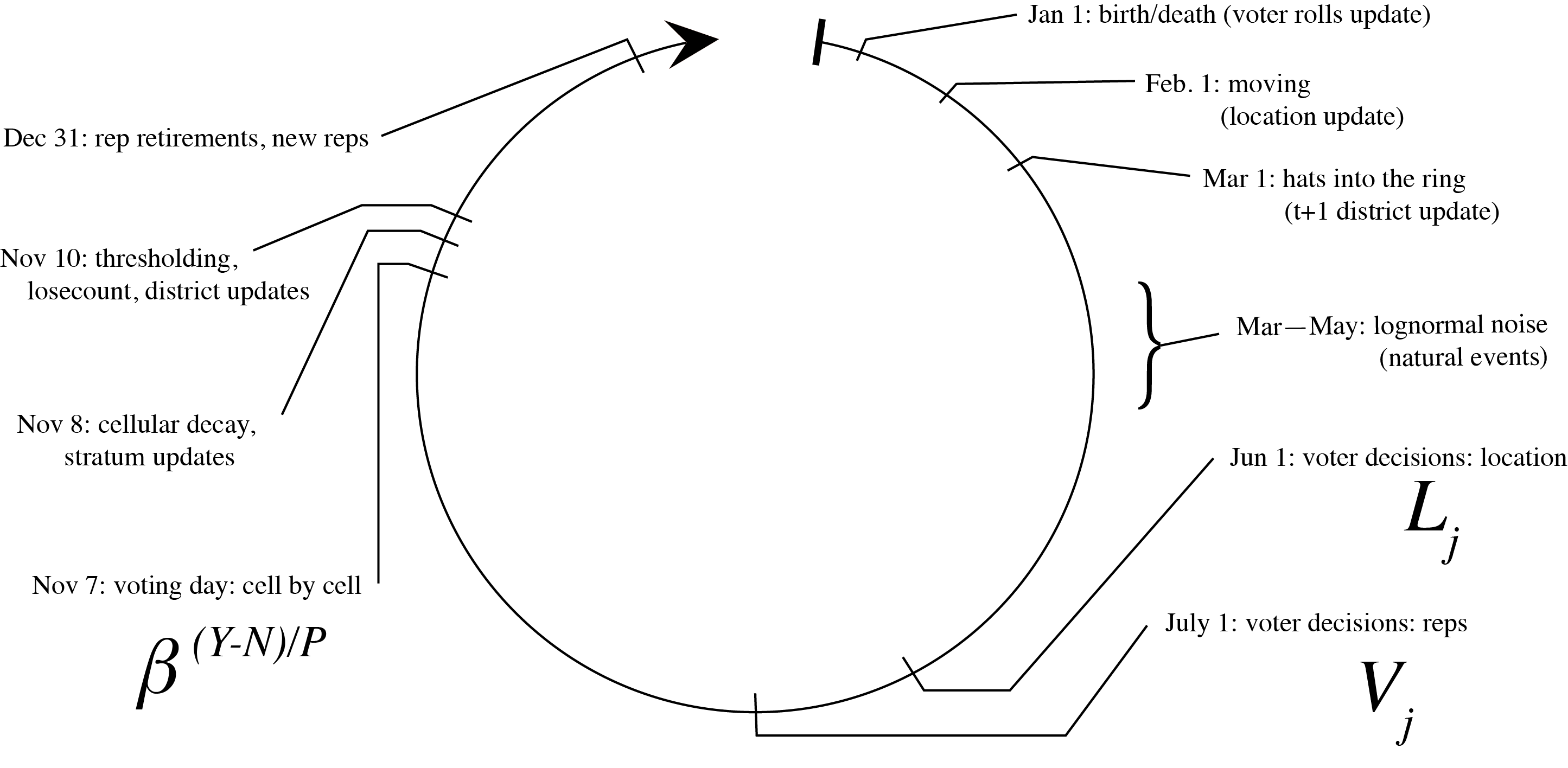}
\end{center}
\caption{CV0 time step}
\label{f.cv0year}
\end{figure}

The sequence of events in a timestep, or election cycle, of the model CV0 is indicated in {\sc Figure} \ref{f.cv0year}. 
For concreteness, we imagine a timestep as the passing of one year, 
though only the order of updates matters. 
The dates are to bring the model in contact with the idea being modeled. 
\enu
	\item {\em Cells.} The map is divided into cells, as described in section \ref{s.cells}. 
The geography (adjacencies, etc.) is not modeled, but can be added in later versions. 
Cells are fixed from one timestep to the next, but their attributes evolve due to slow (small deviation) correlated, lognormal noise. 
Population density of cells for now is governed by voter decision-making alone. 
The number $K$ of cells is held constant. 
	\item {\em Voters.} A voter is essentially a vector of attributes. In particular, voters are not associated to one another in any way. One of these attributes is age, initialized at 0, and another is location, initialized randomly to a cell. Voter age evolves under correlated, lognormal noise. When the voter's age reaches its maximum, currently set at 60, the voter is reset and randomly initialized. (Intuitively, the voter dies and is replaced with a newly ``born'' voter.) The number $N$ of voters is held constant. 
	\item {\em Reps.} As in section \ref{ss.valence}, there is no distinction between incumbents, candidates, and representatives. All are called representatives. Representatives are modeled by {\em reps,} which are again essentially vectors of attributes being updated. Reps are not voters. 
Most of their attributes are fixed, but some, such as ambition, may change or be affected by other attributes, such as momentum. 
The {\em rep population} is the set of all representatives. 
The distinction between those who do and do not hold a seat has little relevancy in the current model, 
because there is no feedback from the reps to the cells or voters. 
In particular there is no entity Congress in the model, no voting or governmental activities by reps that affect either voters or cells. All that is modeled is the affinity and preferences of voters and 
reps, along with their locations (districts, respectively). 
Each rep has a {\em home cell}, the cell where they first ran (or first won a seat, you can say, if they eventually won there). 
The total number $M$ of reps is held constant. 
	\item {\em Moving.} At each timestep the cells on the map are ranked by each voter. 
The general, widely held preference not to move is built into the rankings, so each voter's present location is awarded a generous premium. 
The voter moves if the highest ranked location exceeds his present location. 
In particular, each voter is assumed to have perfect and complete information about cells when decisions about location are made. 
There is no barrier to moving other than the innate preference not to move, which is correlated to other attributes such as income, savings, etc. 
High population density negatively impacts voter location ranking but this varies by voter. 
	\item {\em Running.} 
All reps automatically run for reelection (or election, as the case may be) until they ``retire'' (see below). At every timestep there are two districts associated to each rep: the {\em jump district} and the {\em seat district}. The latter can be null if the rep does not yet hold a seat. 
The {\em size} of any district is the number of cells it contains. 
The seat district is the district that the rep currently represents, from which the rep derives his or her current valence. 
In the model currently, this doesn't have too much effect other than determining the rep's stratum, 
and feeding back into voter location decisions. 
Jump districts are updated early in each cycle using a preference ranking that varies from rep to rep, and is updated during each election cycle. 
Districts that are highly ranked are added to the jump district; the exact number depends on attributes such as ambition, momentum, etc. 
and the rep's stratum (see below). 
	\item {\em Voting.} 
Voters vote Yes or No on the question of whether a jump district is approved, by ordering the reps using a preference score assigned to each rep. These preferences are converted into votes via a neutrality threshold, currently set at 48-52\% favorable. A preference higher than that is converted into a vote of Yes on the rep if the rep is running in the cell, and a vote below that is a No. 
Reps not running in the cell are voted Neutral automatically by all voters in the cell. 
Voting in the model is a frictionless, costless record of the voting population's sentiment. After the location updates, the voter rolls are automatically updated. 
There is no voter registration. 
All voters vote. 
After the voting is calculated, valences are calculated from the votes using the formulas in section \ref{ss.valence}. 
	\item {\em New reps.} 
After each election some reps may retire. 
Their attributes are reset and randomly initialized. New reps have no sitting district, and their jump district is set to the home cell of the retiring rep they replace. 
This is {\em stratum zero}, see below. 
	\item {\em Strata.} 
Strata begin at zero and increase indefinitely. The highest stratum is determined by the value $K$. 
Strata are denoted $S0, S1, S2, \dots$. With each stratum $Si$ is defined a value $s_i$, the {\em size factor} at $Si$, and a value $g_i$, the {\em growth factor} at $Si$. 
The values $g_1, g_2, \dots$ and the values $s_1, s_2, \dots$ increase. 
Each rep has a stratum, which is initialized to zero. A stratum is entered by a rep when a rep's seat district's size exceeds $s_i$. 
Strata determine what occurs when the jump district is updated. 
When a rep is in stratum $Si$, the number of districts the rep may add updating his or her jump district is at most $g_i$. It may be less than that or it may be zero. 
The number cannot be less than zero: the only way a cell can be removed from a rep's district is through decay (negative valence: see section \ref{ss.valence}). 
After the election and the raw tallies are found, valences are calculated, but they are not yet assigned. 
First, the Win Rule is applied to each jump district: if the valence of the rep exceeds the valence of every other rep in the same stratum whose district intersects that rep's district in one or more cells, then the rep is said to have Won, and the rep's seat district is updated to the jump district in the next time step, minus the decay that occured. 
Otherwise the rep is said to have Lost. In that case the rep's losscount is incremented, and the rep's seat district is not modified, but any decay that occured is applied. 
An exception occurs if the rep's jump district is the same as the seat district. In that case, 
the losscount is not incremented. 
After this, all the seat districts have been updated, and the strata are then updated accordingly. If a rep's stratum changes, his losscount is reset. 
	\item {\em Retirement.}
Reps retire if their losscount exceeds a value currently set at 4, if they remain in a given stratum for a number of timesteps currently set at 10, or if they drop by two or more strata in a single time step. 
\Enu

There is a universally distributed penalty/reward each year that models economic outcomes, 
and feeds into the decision making and preference ranking functions. 
However this function is memoriless. More realistic boom/bust patterns are not modeled.

\subsection{Implementation}\label{ss.implementation}
Our modeling was done using an early 2015 MacBook Pro running OS X version 10.12.6 (Sierra). 
We built the model in C++11 using tools from the standard library, the Boost library \cite{BoostLibrary}, and the Armadillo package \cite{ArmadilloCpp}. 
The results and analysis of our simulations will be released in an update of this paper. 
Our source code will be released along with results and analysis.

\section{Conclusion}\label{s.conc}

One of the potential costs of 
any protection against gerrymandering is the cognitive burden on voters. 
By {\em cognitive burden} we refer to the intellectual work that each voter 
has to do in order to participate in the system. 
The limit on cognitive burden 
is a basic reason why representative democracy exists in the first place, since 
without it, 
in the terminology of section \ref{ss.valence}, 
a perfectly good system could assign each representative valence $1 \over N$, where $N$ is the total population, 
and would require ordinary citizens to participate in the Congressional process. This system is known as direct democracy. 
The system of democratic cellular voting outlined in section \ref{s.cells} has a cognitive burden lying somewhere between 
that of direct democracy and that of a familiar republican democracy with a fixed-district plurality voting system.

The mathematical model CV0 introduced in section \ref{s.cv0} is a convenient model, and it is completely defined, i.e., it could be put in practice, theoretically speaking. But it is not difficult to see that the cognitive burden is still too high for it to be a feasible system for a large, living population to use. 
Consider a simple thought experiment. If there are $k$ cells, and each cell is allowed to have one representative, and each representative 
is then allowed to expand by running in any other cell, then it is theoretically possible that all the representatives choose to expand in one cell, say, Cell~$A$. 
If that occurs there will be at least $k$ representatives on the ballot in Cell~$A$ in the next election. 
This is unworkable if a realistic value of $k$ is a thousand, or a million, or ten million. 
Let us say it is not then a {\em feasible} democratic cellular voting system. 
However, democratic cellular voting, as a concept, determines a very broad class of voting systems, so we turn next to the question of how to define systems that 
bring the cognitive burden down to a feasible level. 

As the thought experiment indicates, there are many reasonable ways to lower the theoretical bound 
without restricting choices in ways that seem arbitrary. For example, a representative who appears on a ballot whose base district is on the other side 
of the country would be considered to be ``coming out of nowhere.'' 
The voters living somewhere cannot be expected to evaluate representatives they do not ``know'' because they live in a 
far distant place and the social network is removed in some sense. We can turn the representative away. 
Going like this it is not long before 
we arrive at the conclusion that what is needed next is a continuous model. 

This is a preliminary version of this paper. The results of the simulations are still pending. 
We will report on these simulations when they are completed, in an update to this paper.

\bibliographystyle{abbrv}
\bibliography{mathmain}

\vspace*{10pt}
\noindent {\footnotesize 
\hspace*{18pt}LUCIUS T. SCHOENBAUM \\
\hspace*{18pt}\address{DEPARTMENT OF MATHEMATICS AND STATISTICS\\
\hspace*{18pt}UNIVERSITY OF SOUTH ALABAMA\\
\hspace*{18pt}MOBILE, ALABAMA 36688-0002}\\
\hspace*{18pt}{\it E-mail}: schoenbaum@southalabama.edu\\

}

\clearpage

\end{document}